\documentclass[useAMS,usenatbib]{mn2e}
\usepackage{psfig,aas_macros}
\usepackage[totalwidth=480pt, totalheight=680pt]{geometry}

\newif\ifAMStwofonts

\title[Offset optical line emission in cluster cores]{The Relation Between Line Emission and Brightest Cluster Galaxies in Three Exceptional Clusters: Evidence for Gas Cooling from the ICM}

\author[Hamer et al.]
{\parbox[h]{\textwidth}{
S. L. Hamer$^{1}$\thanks{E-mail:s.l.hamer@durham.ac.uk}, A. C. Edge$^{1}$, A. M. Swinbank$^{1}$, R. J. Wilman$^{2}$,  H. R. Russell$^{3}$, \\ 
A.C. Fabian$^{4}$, J.S. Sanders$^{4}$ \& P. Salom\'e$^{5}$}\\
\vspace*{6pt}\\ 
$^{1}$Institute for Computational Cosmology, Department of Physics, University 
of Durham, South Road, Durham DH1 3LE\\
$^{2}$School of Physics, University of Melbourne, Parkville, Victoria 3010, Australia\\
$^{3}$Department of Physics and Astronomy, University of Waterloo, Waterloo, ON N2L 3G1, Canada \\
$^{4}$Institute of Astronomy, Madingley Road, Cambridge CB3 0HA\\
$^{5}$LERMA, Observatoire de Paris, UMR 8112 du CNRS, 75014, Paris, France\\}
\begin{document}

\date{Accepted TBD; Received TBD; in original form September 2011}

\pagerange{\pageref{firstpage}--\pageref{lastpage}} \pubyear{2011}

\maketitle

\label{firstpage}

\begin{abstract}

There is a strong spatial correlation between brightest cluster
galaxies (BCGs) and the peak density and cooling rate of the intra-cluster medium 
(ICM).  In this paper we combine integral field spectroscopy, CO observations and X-ray 
data to study three exceptional clusters (Abell~1991, Abell~3444 and Ophiuchus)
where there is a physical and dynamical offset between the
BCG and the cooling peak to investigate the connection between the cooling of the 
intracluster medium, the cold gas being deposited and the central galaxy.
We find the majority of the optical line emission is spatially coincident
with the peak in the soft X-rays.  In the case of A1991 we make separate
detections of CO(2-1) emission on the BCG and on the peak of the soft X-ray emission 
suggesting that cooling continues
to occur in the core despite being offset from the BCG.
We conclude that there is a causal link between the lowest
temperature ($<$2\,keV) ICM gas and the molecular gas($\sim$\,30\,K). 
This link is only apparent in systems where a transitory event has decoupled the BCG
from the soft X-ray peak.  We discuss the prospects for identifying more examples of this
rare configuration.

\end{abstract}


\section{Introduction}

Observations of the central regions of massive galaxy clusters show intense X-ray emission suggesting that the ICM is undergoing significant 
radiative cooling \citep{fab81}.  X-ray emitting gas cools at a rate of $t_{cool}$~$\propto$~1/$n$ 
(where $n$ is the gas density), suggesting a 
substantial mass ($\sim$10$^{12}$ M$_{\odot}$) of cool gas should be deposited into the central 
regions and form stars \citep[for a review see][]{fab94}.  In typical cases cooling rates 
of 50--100 M$_{\odot}$yr$^{-1}$ are expected rising as high as $>$500 
M$_{\odot}$yr$^{-1}$ in the most extreme systems, 
although sub-mm observations did not detect this gas in molecular form \citep{ode94}. 
However, through systematic surveys for CO in clusters, 
molecular gas masses of $10^{9-11} M_{\odot}$ have been found \citep{edg01,sc03}.
More recent X-ray observations have revealed much reduced cooling rates ($\sim$1--10 M$_\odot$~yr$^{-1}$) with a sharp truncation 
in cooling at low ($<$~1 keV) X-ray temperatures, suggesting a continuous feedback effect is reheating the gas.
This heavily suppressed cooling has reduced the large discrepancy 
between gas deposition rate and observed molecular gas \citep[see the review by][]{pf06}.

The clusters X-ray luminosity and temperature relate closely to the intrinsic luminosity of the BCG \citep{sch87,sch88,es91b,es91a}. 
This connection between the BCG and cluster properties makes BCGs an important tool for studying clusters.
Since the dense environment at the centre of clusters means that BCG 
experiences a much higher ICM pressure than any other cluster member galaxy 
radiative cooling is at its strongest and
gas cooling from the hot phase  can be accreted onto the BCG, fuelling its evolution. 
However, these regions are also densely populated by other galaxies making the likelihood of interactions 
and mergers much more likely.  Both mergers and gas accretion 
are therefore likely to play important roles in the growth 
and evolution of the BCG and cluster core \citep{tre90,wil06}.

Many BCGs of the most rapidly cooling cluster cores exhibit optical line emission \citep{cav08}.
\citet{cra99} found significant line emission in 32~per cent of a sample of 201 BCGs selected from the BCS X-ray selected sample \citep{ebe98}.
This line emitting gas at 10$^4$ K traces filamentary structures around the BCG \citep{hat05,hat06,mcd11} and 
direct comparison has shown qualitatively similar structures in the 10$^7$ K X-ray \citep{fab08} and 30 K molecular \citep{sal11} gas. 
This structural similarity suggests that the gas phases are linked and as such, studying the line emitting gas can 
give an insight into the physical mechanisms operating within the cluster cores.

\begin{figure}
\psfig{figure=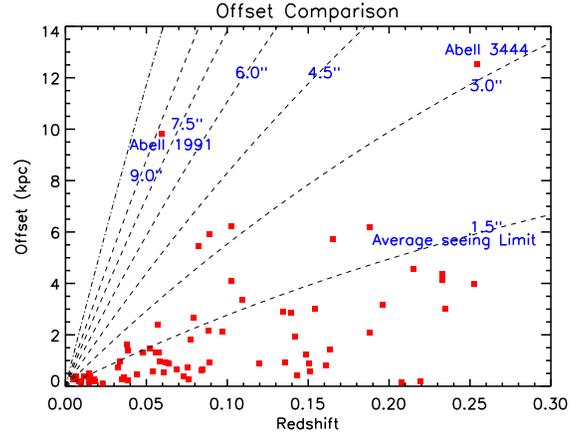,height=6cm,angle=90}
\caption{Observed physical offset between the BCG and the majority of the line emission in our parent sample of 77 BCGs (Hamer {\em et al.} {in preparation}). The dashed lines show constant offsets in 1.5$''$ steps. The dot-dashed line shows the extent of the VIMOS field of view ($\sim$ 12$''$ as the BCG was positioned at the centre of the field of view). The majority of objects fall well below the average seeing limit of 1.5$''$ but A1991 and A3444 stand out with offsets of 9--13\,kpc.}
\label{fig:offsetcomp}
\end{figure}

In order to measure the emission line structure of the gas within a more complete
sample of clusters we have obtained spatially resolved spectroscopy around the 
redshifted H$\alpha$ emission of 77 BCGs
(Hamer {\em et al.} {in preparation}).  Here, we report observations of three extreme 
systems (Abell 1991, Abell 3444 and the Ophiuchus cluster)
from our parent sample which show a significant component ( $>$~50~per cent) of their line 
emission significantly offset from the central BCG. 
A1991 and A3444 are unique from the rest of the sample in that the rest frame optical 
line emission is significantly offset from the BCG.   
Figure \ref{fig:offsetcomp} shows the physical offset between the BCG and the peak 
H$\alpha$ emission for the whole sample of 77 line emitting BCGs (Hamer {\em et al.} 
{in preparation}). It can clearly be seen here 
that the offsets for the majority of the sample fall below the seeing limit of the VIMOS observations (1.5$''$) with A1991 and A3444 having 9-13\,kpc offsets. 
  
We first 
outline our observations and data reduction ($\S$\ref{sec:obs}) followed by a
summary of how we completed our analysis ($\S$\ref{sec:aly}).  In $\S$\ref{sec:res}
we present our results before discussing their implications in $\S$\ref{sec:dis}.
Finally a summary and conclusions are reported in $\S$\ref{sec:sum}.
We assume $\Omega_{m}=$0.27, $\Lambda=$ 0.73 and $H_{o}=$ 71 km s$^{-1}$ Mpc$^{-1}$ throughout.  This corresponds to a physical scale of 1.12\,kpc/$''$ for A1991, 3.91\,kpc/$''$ for A3444 and 0.55\,kpc/$''$ for Ophiuchus.

\begin{figure*}
\psfig{figure=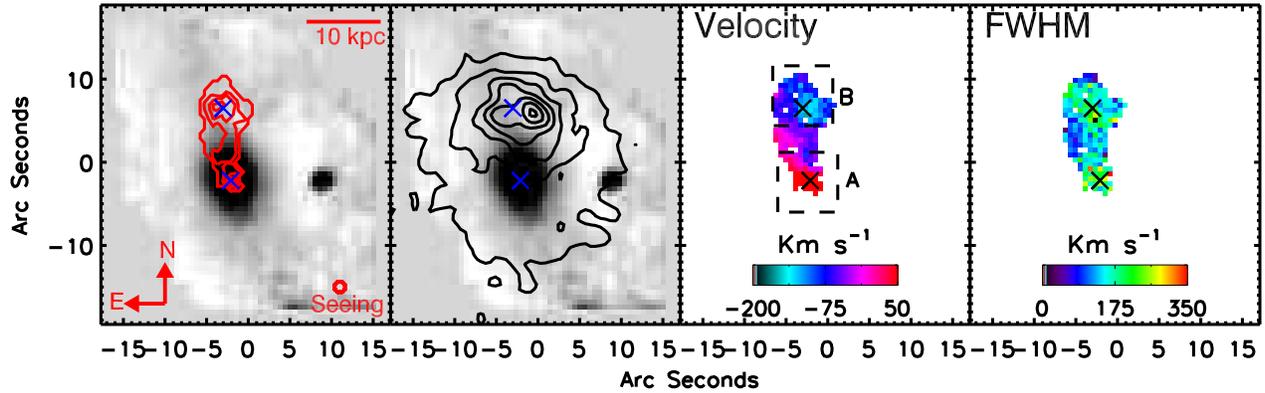,height=6cm}
\caption{{\em Left:} A continuum image ($\lambda$~=~6000-7000~\AA) created from the VIMOS datacube. Overplotted as contours is the 
 H$\alpha$ emission at 6, 12, 18 and 24$\sigma$ (with 1$\sigma$ = 9.4 $\times$ 10$^{-18}$ erg cm$^{-2}$ s$^{-1}$ arcsec$^{-2}$). The H$\alpha$ emission is double peaked with one component on the BCG but extending north 
by 10$''$ where a second component can be seen.
{\em Centre-Left:} The continuum image showing the optical stellar component of the BCG with the X-ray map overlayed as contours.  
The peak in the X-rays is offset by $\sim$ 10$''$ north of the BCG.
{\em Centre-Right: }The H$\alpha$ velocity field for 
A1991, as derived from fits to the H$\alpha$ - [NII] complex. The velocities are given 
relative to a redshift z=0.0591.
{\em Right: } The FWHM of the H$\alpha$ line emission deconvolved for instrumental resolution.
The crosses mark the location of the two H$\alpha$ components on each plot.
The plots are centred at RA = 14:54:31.567 and Dec = 18:38:32.70 (J2000).  There 
is a component of H$\alpha$ emission offset to the north of the BCG by $\sim$\,10$''$ 
which coincides with the position of the peak in the soft X-ray emission.  The 
velocity gradient along this offset shows a dynamical link still exists between the BCG 
and the Peak of ICM cooling.}
\label{fig:offset}
\end{figure*}

\section{Observations and Data Reduction}
\label{sec:obs}
Observations of the BCG in A1991 were taken using the  
Visible Multiobject Spectrograph (VIMOS) instrument on the 8.2m Very Large Telescope 
(VLT) in April of 2008 and May of 2010.
Observations were made at two positions one centred on the BCG and the second $\sim$ 10 $''$ north of the first to cover the peak in X-ray emission.
 At both locations a set of three 600 second exposures were performed
with a pointing dither included between each exposure to account for bad pixels. The HR\_Orange Grism and 
GG435 filter (spectral resolution of R\,$\sim$\,$\Delta\lambda/\lambda$\,$\sim$\,2650 over the wavelength range 5250--7400 \AA) were used to observe H$\alpha $ ($\lambda_{\rm rest} \ 6562.8 \rm$ \AA) at the 
redshift of the cluster of 0.0587.  Observations of A3444 where taken with VIMOS 
in June 2008 using the HR\_Red Grism and GG475 filter (spectral resolution R\,$\sim$\,$\Delta\lambda/\lambda$\,$\sim$\,3100 covering 
the wavelength range 6450--8600 \AA) to observe H$\alpha $ at the cluster redshift of 
0.2533.  The observations were taken in a range of conditions,
with a seeing of $\sim$ 0.7$''$--1.3$''$.  

The raw data were reduced using the {\sl esorex} package. This package performed the basic data 
reduction including bias subtractions, flat fielding and the wavelength and flux calibrations. To subtract the sky we masked any point-like objects to remove any stars within the field. The BCG was then removed from the field by masking all pixels within an isophote of half its peak intensity.  The sky level for each quadrant was then calculated by taking the median value of the remaining pixels at each wavelength increment. This sky spectrum was then subtracted from each pixel in the four quadrants before they were combine into a single datacube.  Finally we median combined the three exposures for each pointing in order to 
eliminate cosmic rays. To mosaic the two positions of A1991 we determine the exact offset 
between the BCG in both observations and combined them to create a cube with a $\sim$ 36$'' \times38''$ field of view.

The Ophiuchus cluster BCG (hereafter Ophiuchus) was observed by \citet{edw09} using Gemini Multi-Object Spectrograph (GMOS) 
on the 8.1m Gemini South Telescope (Gemini).
This object was selected from the literature as it is the best other example of a BCG showing 
significant offset between the BCG, the X-ray peak and the line emission \citep{ma09}.
The observations of Ophiuchus consisted of four 1500 second exposures which were combined to 
give a total integration time of 6000 seconds.  The  R400+r grating (spectral resolution 
R\,$\sim$\,$\Delta\lambda/\lambda$\,$\sim$\,1918 covering the wavelength range 5620--6980 \AA) was used to ensure 
coverage of H$\alpha$ at the cluster redshift of 0.0280 \citep{lah89}.  The seeing during these observations was $\sim$ 0.8$''$ to 1$''$.

The data were reduced using GMOS specific tools in the Gemini IRAF package, 
which included bias subtraction, flat fielding and wavelength calibration. 
Sky subtraction was preformed by isolating the fibres from a secondary IFU situated $\sim$ 1$'$ away from the primary IFU. These fibres were then used to create a sky spectrum which was subtracted from the fibres containing the observations.  Cosmic rays were then rejected by comparing the spectrum from each fibre to that from neighbouring fibres and rejecting any element which had a value exceeding ten times the Poisson noise.  The individual exposures were then combined into a single datacube which covers a region centred on the BCG of $\sim 5'' \times 7''$.

We obtained IRAM 30m data for the CO(1-0) and CO(2-1) lines in A1991 on 21st April 2010.
The observations were performed 
in reasonable conditions ($\tau_{\rm 225GHz}\sim 0.1-0.2$) with the EMIR receiver using a 4~GHz bandwidth
covering each line. Two positions were observed, the BCG and the optical line peak 11$''$ north, 
each for 1 hour duration with wobbler switching with a 90$''$ throw. The observations
reached a noise level of 0.6~mK and 1.3~mK in 44~km~s$^{-1}$ bins (16 and 32~MHz) for CO(1-0) and CO(2-1)
lines respectively. Both the 4~MHz and Wilma backends were used to sample the data but the noise
performance of the 4~MHz backend was considerably better so only the 4~MHz data are presented here.

IRAM 30m observations of CO(1-0) in A3444 were obtained on 29th December 2006,
before the VIMOS IFU observations, in
very good conditions ($\tau_{\rm 225GHz}<0.1$) with the A/B100 receivers with a 400~MHz bandwidth
for 1 hour duration using wobbler switching with a 90$''$ throw. These observations
reached 0.8~mK in 52~km~s$^{-1}$ bins (16~MHz). At the time of writing, we
are aware of no observations of CO in Ophiuchus.

\begin{figure}
\psfig{figure=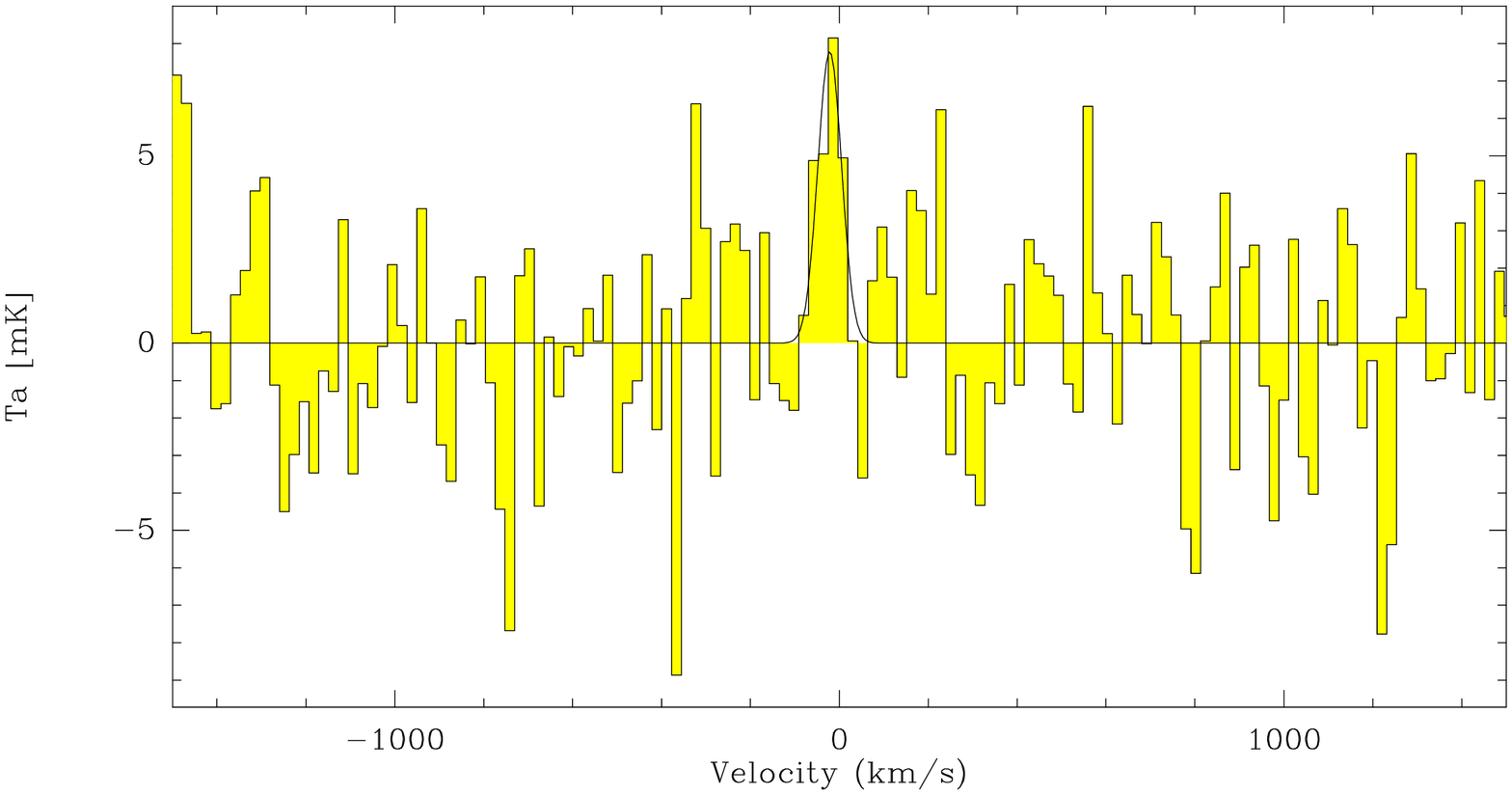,angle=270,height=4.5cm}
\psfig{figure=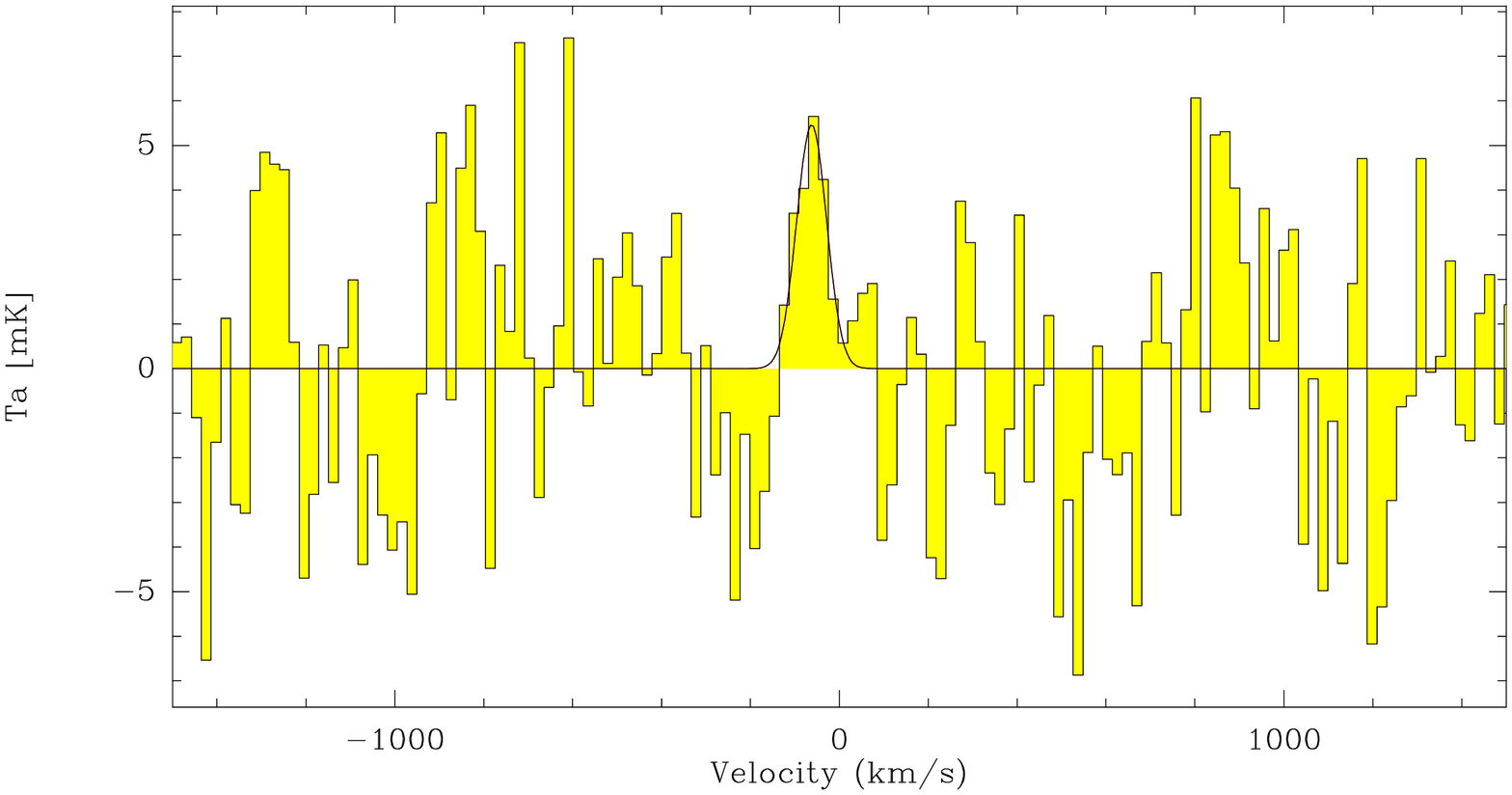,angle=270,height=4.5cm}
\caption{{\it Upper} IRAM 30m EMIR spectrum for A1991 on the
BCG. {\it Lower} IRAM 30m EMIR spectrum for A1991 on the offset
optical line peak. The plots show the CO(2-1) line in antenna temperature scale.  
Significant CO emission is detected at
both the position of the BCG (3.8$\sigma$) and the offset H$\alpha$ emission (3.0$\sigma$) 
and there is a small velocity offset in the line centroid between them.}
\label{fig:co21}
\end{figure}

\section{Analysis}
\label{sec:aly}
We use the integral field spectroscopy to investigate the spatial variation of line emission and kinematics within the clusters.  
We fitted gaussian emission line profiles to the region of the spectra containing
the H$\alpha$ ($\lambda_{\rm rest} \ 6562.8 \rm$ \AA) and [NII] ($\lambda_{\rm rest} \ 6548.1/6583 \rm$ \AA) in each lenslet. We used a $\chi^2$ minimisation procedure and 
allowed the velocity,
 intensity and linewidth of the H$\alpha$ line, as well as the continuum level to vary independently.  The [NII] lines were fixed to the 
same velocity and linewidth as H$\alpha$,
 the [NII] intensity was allowed to vary but the ratio of [NII]6548.1/[NII]6583 was 
fixed to 0.3. The profiles were fitted to each 0.6$''$ pixel and adaptively binned to 
1.8$''$ in regions with lower H$\alpha$ flux. As such the low surface brightness 
emission has a lower resolution  than the brightest regions. 

These fits were accepted as representing the data when they provided an improvement over
a continuum only fit at the 7$\sigma$ significance level and when an acceptable fit was 
found the parameters of the best fit model were stored.  These parameters were then used
to produce maps of H$\alpha$ flux, relative velocity and linewidth(FWHM deconvolved for 
instrumental resolution).  Continuum
images of the region covered by the observations were also produced by taking the 
median of the emission from each lenslet over a region of the spectra containing no
emission lines or sky line residuals (here after referred to as collapsing the cube).

\begin{table*}
\begin{center}
\small
\centerline{\sc Table \ref{tab:fits}.}
\centerline{\sc Properties of the on source and offset components}
\smallskip 
\begin{tabular}{| l || c | c | c | c | c | c |}
\hline
\noalign{\smallskip} 
 Target & Redshift & Velocity & H$\alpha$ Flux & [NII] Flux & Continuum & FWHM \\
 & & (km~s$^{-1}$) & ($10^{-16}$~erg~cm$^{-2}$~s$^{-1}$) & ($10^{-16}$~erg~cm$^{-2}$~s$^{-1}$) & ($10^{-16}$~erg~cm$^{-2}$~s$^{-1} \rm$ \AA$^{-1}$) & (km~s$^{-1}$) \\
\hline
Abell 1991 & & & & & & \\
BCG & 0.05935[3] & 0 $\pm$ 11 & 10.3 $\pm$ 0.7 & 23.7 $\pm$ 0.68 & 6.41 $\pm$ 0.27 & 205 $\pm$ 9  \\
\noalign{\smallskip}
Offset & 0.05893[2] & 124 $\pm$ 10 & 22.5 $\pm$ 1.2 & 35.1 $\pm$ 0.87 & 0.76 $\pm$ 0.41 & 152 $\pm$ 5 \\
\noalign{\smallskip}
Connecting & 0.05910[3] & 71 $\pm$ 11 & 7.48 $\pm$ 0.4 & 10.1 $\pm$ 0.35 & 2.15 $\pm$ 0.19 & 133 $\pm$ 7 \\ 
\hline
Abell 3444 & & & & & & \\
BCG & 0.25583[6] & 0 $\pm$ 25 & 15.5 $\pm$ 0.7 & 8.3 $\pm$ 0.6 & 1.0 $\pm$ 0.2 & 490 $\pm$ 21  \\
\noalign{\smallskip} 
Offset & 0.25550[3] & 100 $\pm$ 19 & 16.5 $\pm$ 0.5 & 7.8 $\pm$ 0.5 & $<$ 0.5 & 273 $\pm$ 9   \\
\noalign{\smallskip} 
Connection & 0.25503[6] & 241 $\pm$ 26 & 10.0 $\pm$ 0.6 & 8.2 $\pm$ 0.6 & 0.4 $\pm$ 0.2 & 450 $\pm$ 26  \\ 
\hline
\end{tabular}
\caption{Results from the fit to the H$\alpha$ [NII] triplet for the total spectrum of each object listed. 
The values in Square brackets are the errors on the last decimal place.  For A1991 the BCG 
emission was defined as coming from a 8$''$ by 7$''$ region centred on the on BCG H$\alpha$ peak, shown in Figure \ref{fig:offset} as region A, the offset emission was defined in a 
similar manner about the offset H$\alpha$ peak (region B in Figure \ref{fig:offset}). 
 Similarly for A3444 the on BCG
emission was defined as coming from a 4$''$ by 5$''$ region centred on the BCG H$\alpha$ peak and the offset was defined as 
an equally sized region about the offset H$\alpha$ peak (region A and B in Figure \ref{fig:offset2} respectively). 
For both objects the connection is defined as the emission which was not in these previous two regions and consists of the low 
surface brightness emission joining the two peaks.}
\label{tab:fits}
\end{center}
\end{table*}

\begin{table*}
\begin{center}
\small
\centerline{\sc Table \ref{tab:coline}.}
\centerline{\sc Properties of the CO(2-1) line in A1991}
\smallskip 
\begin{tabular}{ l  c  c  c  c  c c }
\hline
\noalign{\smallskip}
Target & Significance & CO(2-1) Line Intensity & Velocity & Line width & Peak intensity T$_{mb}$ & H$_2$ gas mass \\
 & Sigma($\sigma$) & (K~km~s$^{-1}$)   & (km~s$^{-1}$) & (km~s$^{-1}$) & (mK)      & ($10^8$~M$_\odot$) \\
\hline
On BCG & 3.8 & 0.52$\pm$0.16 & -22$\pm$10 & 60$\pm$18 & 7.8$\pm$2.2 & 9.9$\pm$3.0 \\
\noalign{\smallskip}
Offset & 3.0 & 0.46$\pm$0.19 & -62$\pm$17 & 79$\pm$35 & 5.5$\pm$1.8 & 8.8$\pm$3.6 \\
\hline
\end{tabular}
\caption{
CO(2-1) line parameters for the IRAM 30m EMIR observations
corrected for beam efficiency. The equivalent analysis for the data for CO(1-0) finds
an upper limit of $<$0.15 K~km~s$^{-1}$ which is consistent
with the CO(2-1) detection given the observed range of line intensities for
these lines in cluster cores (Edge 2001, Salom\'e \& Combes 2003).
}
\label{tab:coline}
\end{center}
\end{table*}

\section{Results}
\label{sec:res}
\subsection{Abell 1991}
In Figure \ref{fig:offset} we show the continuum image of A1991
 (collapsed over $\lambda$~=~6000-7000~\AA) and overlay contours of H$\alpha$ line 
emission.  It is apparent that while there is a component of the 
H$\alpha$ emission centred on the BCG a more extended clump is present 9$''$ north 
lying off the bright stellar component, a projected 
physical offset of $\sim$~10\,kpc at the cluster redshift of 0.0587 \citep{str99}.

Table \ref{tab:fits} shows the values of the variable parameters for 
the minimised fits 
to the emission on the BCG, off the BCG (labelled on Figure \ref{fig:offset} as regions 
A and B respectively) and the emission from the connecting gas.  These 
fits show that the offset clump contains the majority ($\sim$55~per cent) of the H$\alpha$ 
emission within the system.

A1991 also is known to have a significant offset ($\sim$ 10$''$)  between the position 
of the BCG and the X-ray peak \citep{sha04s}. Figure \ref{fig:offset} shows 
that the bright peak in 
the X-ray gas lies roughly 10$''$ (11\,kpc) to the north of the BCG and within 3\,kpc of a 
similar peak in H$\alpha$ emission.  This bright X-ray peak 
corresponds to the lowest temperature gas in the cluster and the region where the gas is cooling
most rapidly. This shows that the majority of the H$\alpha$ emitting gas is related to the most rapidly 
cooling region of the ICM rather than being co-located with the BCG.
Analysis of the {\em Chandra} data indicates that
the pressure in the core of the cluster is relatively uniform between the
BCG and the cooler, offset peak implying that the enhanced density at
the peak is balanced by a lower temperature \citep{sha04s}. 

 In Figure \ref{fig:offset} we show the velocity and linewidth 
maps produced from the fits to the individual lenslets.  We note that the velocity map 
shows a continuous velocity gradient from the position of the BCG along a connecting 
filament to the offset emission.  This suggests that the two components, while physically
separated remain kinematically linked. To determine the velocity difference between the two components
we extracted spectra from an 8$''$ by 7$''$ aperture centred on the BCG and on the offset component
(marked A and B on the velocity map of Figure \ref{fig:offset}).
We find a velocity difference of 124 $\pm$ 10 km s$^{-1}$ between the two components.  
The linewidth (FWHM) of the emission
from the BCG is greater (205 $\pm$ 9 km s$^{-1}$) than that from the offset component
(152 $\pm$ 5 km s$^{-1}$).  

Next, we combine the H$\alpha$ and the CO properties to investigate the relation between 
warm and cold gas within the cluster core.  
By virtue of the good conditions,
the relative sensitivity between the two lines with EMIR strongly favours
the CO(2-1) line. We obtain a significant detection of it in both
positions (Figure \ref{fig:co21}) of 3.8$\sigma$ on the BCG 
and 3.0$\sigma$ on the offset line emission. The
significance of the line detection was calculated using
the variance of the signal away from the line compared
to the line strength. By this measure there are two candidate
lines at -1250 and +930~km s$^{-1}$ in the offset observation
but, if they are real, are unlikely to be related to the BCG.

The separation of the two positions is just smaller than the
IRAM beam at this frequency (10.3$''$ vs 10.9$''$) and the
measured velocity of the emission is offset by more than
half the line width. 
>From the separation and profile of the beam 
we estimate a less than 10 per cent 
contamination between the two pointings. 
Therefore the gas is likely to be spatially
extended on a scale comparable, but not necessarily identical
to, the optical emission lines.
The velocity difference in CO(2-1) between the two positions is -40\,$\pm$\,20\,km\,s$^{-1}$
in the same direction as the shift we observe in H$\alpha$ but 
84\,km\,s$^{-1}$. 
lower. This difference may be related to the bulk flow of the cold molecular gas 
or a different overall distribution with respect to the BCG compared to that traced by the 
optical lines.
The significance of this velocity difference and its potential origin are
beyond the scope of these single dish observations but could be addressed with deeper interferometry
with PdBI or ALMA.

\begin{figure*}
\psfig{figure=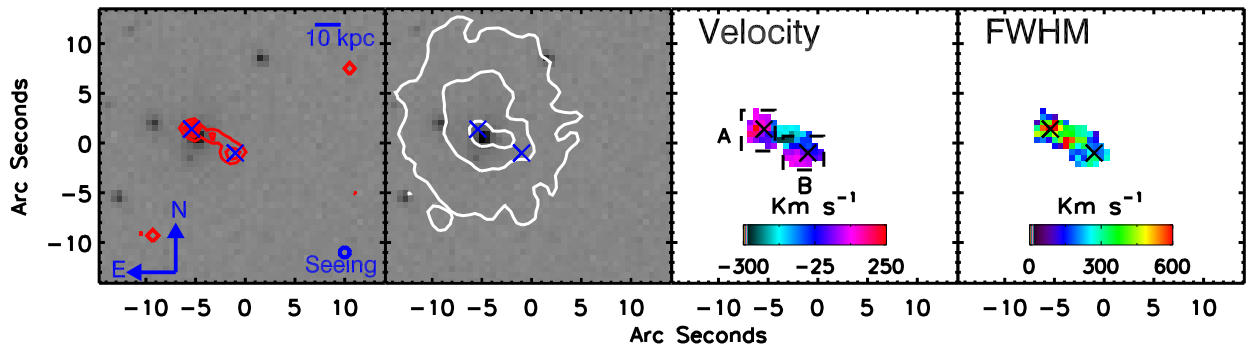,height=5.5cm}
\caption{{\em Left: }This shows a FORS2 R band image of the central region of A3444. Overplotted as contours is the H$\alpha$ line emission at 6,12,18 and 24 $\sigma$ (with 1 $\sigma$ = 
0.021$\times$ 10$^{-16}$ erg cm$^{-2}$ s$^{-1}$).
The H$\alpha$ emission is present towards the centre of the BCG but extends $\sim$ 5$''$ to the southwest where it shows a second peak.
{\em Centre Left: }The FORS2 R band image showing the BCG in A3444 with the {\em Chandra} X-ray map contoured over it.  
While the peak in the X-ray emission lies on the BCG a clear extent can be seen in the south westerly direction in the brightest emission traced by the inner most contour.
{\em Centre Right: }The H$\alpha$ velocity field for A3444, as derived from fits to the H$\alpha$ - [NII] complex. The velocities are given relative to a redshift of z=0.2555.
{\em Right: }The FWHM of the H$\alpha$ line emission deconvolved for instrumental resolution. The crosses mark the location of the two H$\alpha$ components on each plot. The plots are centred at RA=10:23:49.945 Dec=-27:15:25.34 (J2000).  A component of H$\alpha$ emission is offset to the south west of the BCG by $\sim$\,5$''$. This component matches the direction of an extent in the soft X-ray emission.}
\label{fig:offset2}
\end{figure*}

The detection of CO 10$''$ north of the BCG in A1991 makes this the fourth
system where extended CO has been found (e.g. NGC1275 \citet{sal06s}, A1795 \citet{sc04} and RXJ0821$+$07 \citet{ef03}). 
In all of these systems the extended CO 
coincides with extended H$\alpha$ emission.
Within the system as a whole the implied total molecular gas masses of $\approx1.9 \times 10^9$ M$_\odot$ for A1991 is consistent
with the correlation of molecular gas mass and optical line luminosity in \citet{edg01}
and with the star formation rate inferred from the MIR continuum \citep{ode08s}.
However, current observations cannot spatially resolve whether the star formation in the 
offset gas follows the correlation found for gas within the BCG \citep{ode08s}.

\subsection{Abell 3444}
A3444 also shows a substantial factor 
($\sim$63~per cent) of optical line emission offset from the BCG.  In Figure 
\ref{fig:offset2} we show a FORS2 $R$ band image of the central region of the cluster
with contours from H$\alpha$ line emission. A clear extension in the line emission
can be seen to the southwest where a secondary peak is apparent. This component is
offset from the BCG by $\sim$~3.2$''$  or  $\sim$~12.5\,kpc at 
the cluster redshift of 0.2533 \citep{str99}.  In this cluster the
line emission form the BCG and offset location are consistent within errors at $\sim$ 40\%
of the total line flux each (Table \ref{tab:fits}).

\begin{figure}
\psfig{figure=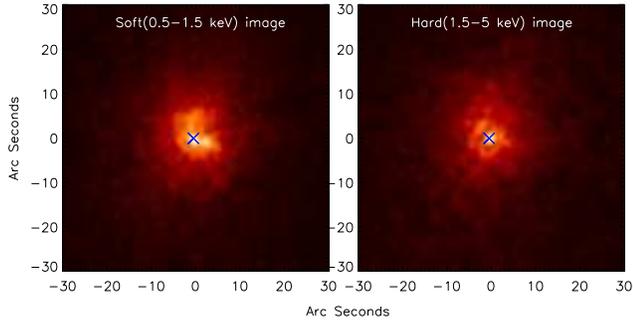,angle=90,height=5cm}
\caption{{\it Left} shows a soft (0.5-1.5 keV) X-ray image of the centre of A3444. A bright peak can be seen at the centre 
of the image which matches the position of the BCG.  A secondary lower intensity peak can be seen to the west-southwest of 
the central peak offset by $\sim$2-3 arcseconds. {\it Right} shows the hard (1.5-5 keV) image of the A3444 cluster core.  
A bright peak can clearly be seen at the centre of this image matching the position of the primary peak in the soft band.
The crosses mark the location of the BCG on each image.
The presence of this unresolved, hard emission on the central peak suggests this emission may be from an AGN component in the BCG.}
\label{fig:xrycomp}
\end{figure}

The X-ray observations of this system show a weak but significant point like source (L$_{x}$(0.5-10 keV) $\sim$~4 $\times$ 
10$^{43}$ ergs s$^{-1}$ in a 1.5$''$ aperture and a power law spectrum) coincident with the BCG
that has a relatively hard spectrum compared to the cluster emission
 (Figure \ref{fig:xrycomp}) suggesting the presence of an AGN.  While the point source is 
centred on the BCG (Figure \ref{fig:offset2}), the soft X-ray 
emission shows an extension in the same direction as the 
offset optical line emission (Figure \ref{fig:offset2}) although,
like A1991, it is not exactly coincident with the offset peak in the line emission. 

To determine the temperature of the gas in at the location of the BCG we fit a mekal 
thermal plasma model \citep{mew85,mew86} to the 0.5--5.0\,keV emission from a 1.5$''$ 
aperture centred on the BCG.  This found the temperature of the gas to be $>$ 7.5 keV
which is high for a plasma model suggesting a 
contribution to the hard X-ray emission from an AGN component.
The presence of an AGN 
in the BCG would produce increased X-ray counts which could shift the apparent centre
of the cluster emission.  As such the AGN contribution must be accounted for when 
determining the true centre of ICM cooling.

To subtract the AGN emission from the soft X-ray emission on the BCG (and find the true level of the extended cluster emission) 
we determine the normalisation of the point source in the 2.0--5.0\,keV band and apply this to the emission from the 0.5--1.5\,keV band.
To do this we first determine the properties of the cluster 2.0--5.0 keV emission from 
an annulus surrounding the point source.  We then fit an absorbed power law plus an 
absorbed mekal plasma model to the point source 2.0--5.0\,keV emission with the 
properties of the mekal component fixed to those from the surrounding annulus.
Finally we apply the same model to the point source 0.5--1.5\,keV emission with the
normalisation and photon index of the power law set to the values determine from 
the Hard X-rays and calculate the flux of the two components.

The total flux from the point source in the 0.5--1.5\,keV band is 2.45$^{+0.24}_{-0.20}$$\times$10$^{-14}$ erg cm$^{-2}$ s$^{-1}$, of which 
0.55$^{+0.16}_{-0.13}$$\times$10$^{-14}$ erg cm$^{-2}$ s$^{-1}$ comes from the AGN contribution.  We find that the cluster emission  under 
the point source is 1.90$^{+0.18}_{-0.15}$$\times$10$^{-14}$ erg cm$^{-2}$ s$^{-1}$ while the extension to the southwest has a flux of 
2.4$^{+0.20}_{-0.12}$$\times$10$^{-14}$ erg cm$^{-2}$ s$^{-1}$ suggesting that it represents the peak in the soft X-ray emission.
Analysis of the {\em Chandra} data also shows that the pressure in the core
of A3444 is relatively constant when the AGN emission is excised. Therefore, A3444 shows very
similar X-ray properties to A1991 with an offset peak in soft X-rays but a constant 
pressure core on 50--100\,kpc scales.
In the case of A3444, where the AGN in the BCG contributes significantly to the X-ray emission,
the offset position in the soft X-rays represents the region of maximum
cooling of the ICM. 


Within this system the BCG and offset component are separated by a velocity
of 100 $\pm$ 19 km s$^{-1}$.  We also note that the connecting emission has a higher 
velocity offset (241 $\pm$ 26 km s$^{-1}$) than the offset peak. The linewidths 
in this system are higher in the BCG (FWHM= 490 $\pm$ 21 km s$^{-1}$) than in the 
offset component (FWHM= 273 $\pm$ 9 km s$^{-1}$)  suggesting the
gas is more disturbed within the galaxy or that it is the combination
of several, related gas clouds.
We also note that the lenslet
from the centre of the BCG has FWHM $>$~650 km s$^{-1}$
suggesting possible contribution from the AGN detected in the X-ray imaging, that can be 
seen in the linewidth map in Figure \ref{fig:offset2}.

In A3444, the smaller angular extent of the offset means that only a global measure
of the cold molecular gas can be obtained from our IRAM observations
in this system and its higher
redshift restricts us to just CO(1-0). Our observation provides a 
marginal (2.0$\sigma$) detection of 0.33$\pm$0.11~K~km~s$^{-1}$ and a width of
172$\pm$69~km~s$^{-1}$ at the velocity of the BCG. However, the
restricted bandwidth of the A/B100 receiver means that a broader
line could be present but is lost in the baseline subtraction.
The implied total molecular gas masses of $\approx1.7 \times 10^{10}$ M$_\odot$ for A3444 
is consistent
with the correlation of molecular gas mass and optical line luminosity in \citet{edg01}.

\subsection{Ophiuchus}
\begin{figure}
\psfig{figure=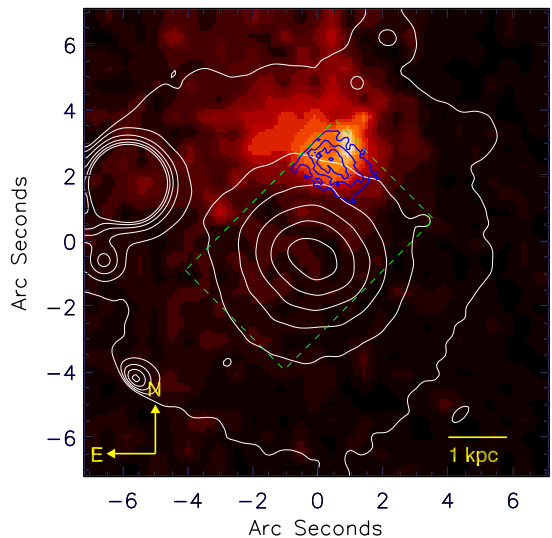,angle=90,height=8cm}
\caption{The colour scale shows the X-ray emission observed by Chandra with contours from the continuum emission from the {\em GMOS} R-band acquisition image of the Ophiuchus BCG. Blue contours show the position of the H$\alpha$ line emission at 6, 12, 18 and 24 $\sigma$. The H$\alpha$ line emission is offset to the north of the BCG. The position of the offset H$\alpha$ emission lies near the peak in the X-ray surface brightness.  It should be noted that the north-east and north-west edges of the line emission coincide with the edges of the {\it GMOS} field of view (shown as dashed green lines) so these contours may not represent the true extent of the H$\alpha$.}
\label{fig:ophu}
\end{figure}

\citet{edw09} studied the properties of the Ophiuchus cluster with the GMOS IFU and 
found line emission with a substantial ($\sim$ 4$''$) offset from the BCG 
(Figure \ref{fig:ophu}).  \citet{edw09} attributed this offset line emission to another galaxy
and not to the BCG itself but given the similarity with respect to the X-ray emission to 
A1991 and A3444 we reconsider this system.
In our reanalysis of these data we also find that 
the line emission is offset from the BCG by $\sim$ 4$''$ which, at a redshift of z=0.028,
corresponds to a projected separation of $\sim 2.2$\,kpc. 
The line emission within Ophiuchus shows no component on the BCG with 100~per cent of the 
H$\alpha$ emission coming from the offset component.

\citet{mil10} also analyse {\em Chandra} X-ray data and find that the X-ray peak in the Ophiuchus cluster is offset from the BCG 
by $\sim$ 4$''$, a similar offset to that seen in the line emission.  In Figure 
\ref{fig:ophu} we compare the position of the X-rays and 
the offset line emission.  We note that the {\it GMOS} field of view 
(shown by the 
dashed green line) ends along the north-east and north-west edges of the observed line 
emission but that otherwise the line emission lies towards the peak in X-ray surface 
brightness. While our analysis of these data find results consistent with those of 
\citet{edw09} the alignment of the offset H$\alpha$ and X-ray emission suggests a 
situation similar to that in A1991 and A3444.

\begin{figure}
\psfig{figure=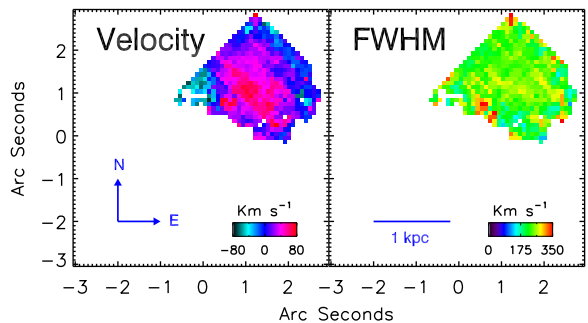,height=5cm}
\caption{The H$\alpha$ velocity field ({\em Left}) and FWHM linewidth ({\em Right}) for 
Ophiuchus, as derived from fits to the H$\alpha$ - [NII] complex. The velocities are given 
relative to a redshift z=0.0304.  The velocity and linewidths are simple and show little variation although tentative evidence for for a velocity structure is apparent to the east of the object.}
\label{fig:kineophu}
\end{figure}

In this system, there is sufficient S/N in the continuum that we can determine the stellar velocity. 
We extracted a spectrum of the BCG and used a penalised pixel-fitting method \citep{cap04} to fit the 
Sodium D absorption from a catalogue of stellar templates.
Comparison of the H$\alpha$ line with the Sodium D($\lambda$ 5889,5895) stellar absorption find that the
line emitting gas is offset from the BCG by 603 $\pm$ 28 km s$^{-1}$.
The velocity
and line width maps are shown in Figure \ref{fig:kineophu}, the kinematics of the gas 
in this system are simple although some tentative evidence for a velocity structure
is apparent to the east of the object (at the edge of the GMOS field of view).  
Other than this the velocity of the gas shows little variation 
($\sigma$ = 31 km s$^{-1}$ or half the velocity resolution of $\sim$~60 km 
s$^{-1}$) and the linewidth is uniform across the detected line emission (FWHM = 235 $\pm$ 38 km 
s$^{-1}$). Future, wider field IFU observations are required to map the full extent
and velocity structure of the line emission.

\subsection{Time scales}

The conclusion we draw from our extensive optical IFU sample (Hamer {\em et al.} {in preparation})
is that the fraction of systems in which there is an offset between the majority of the
optical line emission and the central galaxy of more than 8\,kpc is relatively small ($\approx 3$ in 77). 
In the narrow-band imaging survey of 27 groups and clusters \citep{mcd11}, A1991
is the only object with a second component of the optical line emission 
that is offset from the BCG. 
Similarly, from X-ray observations of 24 line emitting clusters from the LoCuSS sample 
\citep{san09} only four
are offset from the X-ray centroid by more than 8\,kpc (i.e. 15~per cent) 
(but none are offset by more than 15\,kpc). 


We can use the statistics of offset line emission of our sample to estimate the lifetime of such offsets.  
Assuming that all clusters with a line emitting BCG may experience an offset at some stage,
the fraction of BCGs showing this offset multiplied by the look-back time of the sample ($\sim$1~Gyr at the median 
redshift of our sample) gives an estimate on the lifetime of $\sim$40~Myr (3/77$\times$1~Gyr).
With IFU observations, it is also possible to estimate the lifetime of the offset using the kinematics of the gas. 
Assuming the offset is orientated to the line of sight at an angle $\theta$, the actual separation of the BCG 
and X-ray peak is $D = D'/sin \theta$, where D is the actual separation and D' is the projected separation that we observe.  
Similarly the projected and actual velocity difference is given by $V = V'/cos \theta$.  We can determine the 
timescale of this separation to be $T = D'/V' \times cos \theta/sin \theta$.

Assuming that the likelihood of any orientation is uniform over the area
a hemisphere, then the median orientation angle will correspond to the
angle that splits the hemisphere into two equal areas. This implies that 2$\pi$r$^3$~cos$\theta_{2}$ = $\pi$r$^3$
and hence $\theta$=cos$^{-1}$~0.5~$\sim$~60$^{\circ}$.



A1991 has the most well constrained parameters of our three objects
and therefore by setting a limit on the offset velocity and distance we can limit the range of 
angles over which we expect the offset to be orientated. We assume the maximum offset 
velocity to be 500~km~s$^{-1}$ and the offset distance to be no more than 20\,kpc.  These 
values correspond to an orientation ranging from $\sim$30$^{\circ}$--75$^{\circ}$.
By applying the above equation with appropriate orientation angles we find that its most likely lifetime is 45~Myr
with a possible range of 20--135~Myrs.
These values are consistent with those calculated from the sample statistics suggesting 
that the lifetime of such offsets is a few tens to a hundred Myrs.

At a mass deposition rate of 10--20\,M$_{\odot}$\,yr$^{-1}$ \citep{sha04s} 2\,$\times$\,10$^7$--2.5\,$\times$\,10$^9$ M$_{\odot}$ could condense from the ICM within 20--135 Myr.  With our most likely timescale of 45 Myr this corresponds to 5\,$\times$\,10$^8$--10$^9$ M$_{\odot}$ so the expected mass of gas is comparable to our detected CO gas mass.  It should be noted that the lower mass deposition rate represents gas centred on the BCG which is at a higher temperature and lower density than that at the offset position so should be considered a lower estimate.

\section{Discussion}
\label{sec:dis}

The most striking result from our observations is that the optical
line emission is emitted predominantly from the region with the highest
soft X-ray surface brightness and not the BCG. 
In the vast majority
of clusters the peak in the X-ray emission is very close to the BCG
 \citep{per98,hud10,san09} so determining whether the optical lines
are primarily related to the cooling of the ICM or the BCG is
not possible. The direct relationship 
between the core X-ray properties (e.g. cooling time or entropy)
and the optical line emission in any BCG has been established \citep{raf08,cav08}.
However, only in the very small minority of clusters
where the `sloshing' of the ICM resulting from major or minor cluster mergers \citep{asc06,jon10,mil10} 
or an episode of strong AGN activity
temporarily results in the BCG not residing
at the  peak of the gas cooling is there an opportunity to make the direct
test of whether the optical line emission is related primarily
with the cooling gas or the BCG itself.

The presence of cold molecular gas in the cores of clusters
can be explained through a number of mechanisms: mergers with 
gas-rich cluster galaxies or gas cooling
from the intracluster medium. The gas mass observed in
BCGs (10$^{9-11.5}$M$_\odot$) \citep{edg01,sc03}
is too large to be explained by galaxy mergers or stellar mass loss but is
consistent with heavily suppressed cooling from the ICM \citep{pet03,pf06,snd10}.

Given the close correlation
between optical lines and CO found in other cluster cores the detection of ionised gas at the location of the X-ray peak suggests
the presence of a reservoir of molecular gas offset from the BCG. In A1991 and A3444 we directly
detect the presence of molecular gas in the form of CO.  In one system, A1991 we also detect
CO at the offset position, with a molecular gas mass of $\sim$10$^{9}$ M$_{\odot}$.
The presence of such a high mass of molecular gas at the offset 
position clearly suggests that cooling is occurring despite the BCG and cluster core being separated.

Using the statistics from our sample we estimate that
any offset of greater than 8\,kpc can only remain for around 10$^8$yrs.
This is consistent with the dynamical timescale of the core
implying that these ``excursions'' of the BCG from the X-ray peak
are short-lived.
If the molecular 
gas at the offset position is being produced in situ then it would have
to cool from the ICM within this time frame.  Within the central regions
of the cluster core X-ray observations predict typical cooling rates of the order
$\sim$ 1-10 M$_{\odot}$ yr$^{-1}$ \citep{pf06}, at this rate of cooling the cluster
core could deposit between 10$^{8}$ and 10$^{9}$M$_{\odot}$ of cold
molecular gas in the $\sim$ 10$^8$ yrs which it is not coincident with the BCG.  
As such, the offset molecular gas observed  in A1991 is consistent with the 
upper band of cooling from the ICM of 25~M$_{\odot}$ yr$^{-1}$ \citep{sha04s}.

Although A3444 and Ophiuchus do not have separate detections of cold molecular gas at the offset position, 
their similarities to A1991 suggest that it may be present.   Due to the smaller angular 
offset between BCG and X-ray peak in A3444 and Ophiuchus, resolving 
the molecular gas is not possible with single dish observations.  
However, with the start of ALMA operations there is
now the possibility to directly map the molecular gas in such systems allowing
the separate components of the gas to be resolved.

An alternative explanation for the morphology of the extended molecular and ionised gas is ram pressure
stripping of gas related to the BCG when a merger induces large motion of
the ICM relative to the BCG. This requires an efficient mechanism to entrain
the dense molecular gas into the much more tenuous ICM. 
The effect of magnetic 
fields could enhance this stripping but would have to be very effective to
explain the acceleration observed in Ophiuchus. The observations presented here
cannot rule out ram pressure stripping. However, future mm-interferometry should be
able to map the cold molecular gas and allow us to distinguish between these mechanisms.
If ram pressure stripping is responsible we would expect the molecular gas to be stripped
less effectively than the optical line-emitting gas, while in the case of local cooling
we would expect a similar morphology between the two phases.

\subsection{Similar objects}
The offset between the BCG and X-ray peak in clusters is not unique to these
objects, both Abell 1644 \citep{jon10} and RXJ1347.5-1145 \citep{jon11}
have a small offset and
show evidence of sloshing similar to that seen here. \citet{mar03} also
find that two thirds of clusters classified as cool cores by \citep{per98}
show evidence of cold fronts associated with sloshing.  
An offset along the line of sight is not always obvious suggesting the existence
of more objects in which the BCG is dissociated from the cluster core but cannot 
be detected by observations due to projection effects.

There are also a number of other systems where 
the peak of the line emission is coincident with the BCG
but have less intense offset line emission in the form of extended envelopes and 
filaments. A1795 \citep{cra05} and RXJ0821$+$07 \citep{bay02}
both show very extended optical emission that follows the highest
X-ray emissivity in the cluster and also has associated
cold molecular gas emission \citep{sc04,ef03}.
Most recently, \citet{can11s} present evidence for
a filament of optical line emission extending from the BCG in
A2146 which is undergoing a major merger where the peak in the
X-ray emission is offset from the BCG by 37~kpc. The IFU data
presented in \citet{can11s} doesn't cover the X-ray peak 
so it can't be established if A2146 is similar to the three clusters
presented here but it illustrates that a significant merger is
required to dissociate the BCG from the X-ray peak. In
a major merger the disruption in the core will be 
significant. \citet{rus10} show that the pressure
profile in the core of A2146 is very asymmetric,
unlike that in A1991 which is dynamically more quiescent.
Therefore, the process of cluster-cluster merging, and the longer lived
sloshing that it induces, may be required to produce the brief decoupling of the
peak in the intracluster medium density from the BCG observed
in all of these systems.

Another viable mechanism for inducing bulk motion of the ICM
in a cluster core is the inflation of cavities due to AGN
activity in the BCG. \citet{bla11} show a close
correspondance between filaments of high X-ray surface
brightness and optical line emission in A2052 which has an extensive
cavity network. The compressed gas in these filaments
has a short cooling time so the optical line
emission may also be related to local cooling. The systems
presented in this paper do not resemble A2052 or
exhibit the large cavities required to generate the 
observed offsets, so AGN feedback is unlikely to
explain the lop-sided offsets we observe in A1991, A3444 and Ophiuchus. However,
the small scale correlation between strongest X-ray 
cooling and optical line emission suggests a common
mechanism behind them.

The existence of cold gas that is `orphaned' in the cores
of clusters may have a number of implications for more
distant systems. For instance, in the high redshift radio
galaxy TXS0828+193 ($z=2.6$), \citet{nes09} find a cloud
of cold molecular gas emitting at CO(3-2) that has no bright
optical or near-infrared counterpart that is 80\,kpc from the
radio galaxy itself. They speculate that this may be
analogous to low redshift clusters but note that the
gas tends to be associated with the central galaxy.
While the total gas mass and physical offset are
substantially larger in TXS0828+193 than in A1991 or A3444, the
similarity between the systems may point to a common
mechanism.

\section{Summary}
\label{sec:sum}
Within the cores of most clusters there is a clear spatial coincidence between the
position of the BCG and the peak of the ICM density. However there are a
few clusters in which the BCG and cluster core are found to be decoupled. 
In such systems, it is possible to disentangle the effects of the cluster
core and the BCG on the ICM and study each individually. Here we study
three such systems selected from a parent sample of 77 massive X-ray
selected clusters where we find that the optical, millimetre and X-ray observations 
of the cores show extended optical line emission and cold molecular 
gas directly related to the peak of the cooling of the intracluster medium and not 
the central galaxy. These three systems are the only clusters from our sample
to show a significant offset between the BCG and cluster core suggesting that such a phenomenon
is rare and transient.  From our sample we find that a significant offset occurs in only
2--3~per cent of systems and that they likely have lifetimes lasting of the order of 20--100~Myr. 
We note however that it is possible for systems to have an offset along
the line of sight which we would not detect so this fraction may be slightly higher and should be considered 
a lower limit.

The rarity of such significant offsets between the BCG, X-ray peak and optical line emission
points to a large event in cluster evolution, a major cluster merger or 
possibly a powerful AGN outburst.
Mergers are predicted to drive shocks through the core of the ICM at early times
and cause sloshing as a later effect \citep{asc06,jon10,mil10,zuh10,zuh11} both of which are capable 
of separating the BCG from the cluster core.
X-ray observations of A1991 suggest the passage of a cluster-wide shock  through the
core of the cluster and similar observations of Ophiuchus show sloshing of the ICM
suggesting the possibility of a recent merger. Whatever the reason for the separation
the gas cooling at the X-ray peak will continue and cooled gas will be deposited away from the BCG.
This provides a unique opportunity to directly constrain the process of gas cooling without the
presence of a BCG.

\section*{Acknowledgements}


SLH acknowledges the support received from an STFC 
studentship. AMS acknowledges an STFC Advanced Fellowship.  HRR acknowledges generous financial support from the Canadian Space Agency Space Science Enhancement Program.  ACF thanks the Royal Society for support.

Based on observations made with ESO Telescopes at the La Silla or Paranal Observatories under programme ID 081.A-0422 and 385.A-0955.

Based on observations carried out with the IRAM 30m Telescope. IRAM is supported by INSU/CNRS (France), MPG (Germany) and IGN (Spain)

Based on observations obtained at the Gemini Observatory, which is operated by the 
Association of Universities for Research in Astronomy, Inc., under a cooperative agreement 
with the NSF on behalf of the Gemini partnership: the National Science Foundation (United 
States), the Science and Technology Facilities Council (United Kingdom), the 
National Research Council (Canada), CONICYT (Chile), the Australian Research Council (Australia), 
Ministério da Ciência, Tecnologia e Inovação (Brazil) 
and Ministerio de Ciencia, Tecnología e Innovación Productiva (Argentina).

\bibliographystyle{mn2e}
\bibliography{bib}



\label{lastpage}

\end{document}